\documentclass[aps,prb,preprint]{revtex4}

\usepackage{graphicx}

\begin{document}

\author {W. J. Mullin}
\author{G. Blaylock} 
\affiliation{Department of Physics, University of Massachusetts, Amherst,
Massachusetts 01003}
\title{Quantum Statistics: Is there an effective fermion repulsion or boson
attraction?}

\begin{abstract}
Physicists often claim that there is an effective repulsion between
fermions, implied by the Pauli principle, and a corresponding effective
attraction between bosons. We examine the origins of such exchange force ideas,
the validity for them, and the areas where they are highly misleading. We
propose that future explanations of quantum statistics should
avoid the idea of a effective force completely and replace it with more
appropriate physical insights, some of which are suggested here.
\end{abstract}

\maketitle

\section{INTRODUCTION}
\label{sec:Intro}
The Pauli principle states that no two fermions can have the same quantum
numbers. The origin of this law is the required antisymmetry of the
multi-fermion wavefunction. Most physicists have heard or read a shorthand
way of expressing the Pauli principle, which says something analogous to
fermions being ``antisocial'' and bosons ``gregarious.'' Quite often this
intuitive approach involves the statement that there is an effective
repulsion between two fermions, sometimes called an ``exchange force,''
that keeps them spacially separated. We inquire into the validity of this
heuristic point of view and find that the suggestion of an effective
repulsion between fermions or an attraction between bosons is actually quite
a dangerous concept, especially for beginning students, since it often
leads to an inaccurate physical interpretation and sometimes to incorrect
results. We argue that the effective interaction interpretation of the Pauli
principle (or Bose principle) should almost always be replaced by some
alternate physical interpretation that better reveals the true physics.

Physics comes in two parts: the precise mathematical formulation of the
laws, and the conceptual interpretation of the mathematics. 
David Layzer says,\cite{Layzer} ``There
is a peculiar synergy between mathematics and ordinary language.\ldots The
two modes of discourse [words and symbols] stimulate and reinforce one
another. Without adequate verbal support, the formulas and diagrams tend to
lose their meaning; without formulas and diagrams, the words and phrases
refuse to take on new meanings.''
Interpreting the meaning of wavefunction symmetry or
antisymmetry in a simple pedagogical
representation is thus vitally important.
However, if those words actually convey the wrong meaning of the
mathematics, they must be replaced by more accurate words. We feel this is
the situation with the heuristic ``effective repulsion'' for fermions
or ``effective attraction'' for bosons, or ``exchange force'' generally.

One can demonstrate there is no real force due to Fermi/Bose 
symmetries by examining a time-dependent wave packet for two 
noninteracting spinless fermions.  Consider the antisymmetric wave 
function for one-dimensional Gaussian wave packets, each satisfying 
the Schr\"{o}dinger equation, and moving towards each other:
\begin{eqnarray}
\psi (x_{1},x_{2},t) &=&C\left\{f(x_{1},x_{2}) \exp \left[ -\alpha (x_{1}-vt+a)^{2}-\beta
\left( x_{2}+vt-a\right) ^{2}\right] \right.   \nonumber \\
&&\left. -f(x_{2},x_{1})\exp \left[ -\alpha (x_{2}-vt+a)^{2}-\beta \left(
x_{1}+vt-a\right) ^{2}\right] \right\}, 
\end{eqnarray}
where $x_{1}$ and $x_{2}$ are the particle coordinates, 
$f(x_{1},x_{2})=\exp\left[imv(x_{1}-x_{2})/\hbar\right]$, $C$ is a 
time-dependent factor, and the packet width parameters $\alpha $ and 
$\beta $ are unequal.  In reality, each single-particle packet 
will spread with time, but we 
assume that the spreading is negligible over the short time that we 
consider the system.  At $t=0$ the $\alpha $-packet is peaked at $-a$ 
and moving to the right with velocity $v$, while the $\beta $-packet 
is peaked at \thinspace $+a$ and traveling to the left with the same 
velocity.  Of course we cannot identify which particle is in which 
packet since they are indistinguishable, and each has a probability of 
being in each packet.  At $t=0$ the packets are assumed well separated 
with little overlap.

At $t=a/v,$ the wave function becomes
\begin{equation}
\psi (x_{1},x_{2},t)=C\left\{ f(x_{1},x_{2})\exp \left[ -\alpha (x_{1})^{2}-\beta \left(
x_{2}\right) ^{2}\right] -f(x_{2},x_{1})\exp \left[ -\alpha (x_{2})^{2}-\beta \left(
x_{1}\right) ^{2}\right] \right\}, 
\end{equation}
and the direct and exchange parts have maximal overlap. The wave function
clearly vanishes at $x_{1}=x_{2}$ (as it does at all times). At the time $%
t=2a/v,$ the packets have passed through one another and overlap very little
again: 
\begin{eqnarray}
\psi (x_{1},x_{2},t) &=&C\left\{f(x_{1},x_{2}) \exp \left[ -\alpha (x_{1}-a)^{2}-\beta
\left( x_{2}+a\right) ^{2}\right] \right.   \nonumber \\
&&\left. -f(x_{2},x_{1})\exp \left[ -\alpha (x_{2}-a)^{2}-\beta \left( x_{1}+a\right)
^{2}\right] \right\}. 
\end{eqnarray}
Now the $\alpha $-packet is peaked at \thinspace $+a,$ but still moving to
the right and the $\beta $-packet is peaked at $-a$ and still moving to the
left. The packets have moved through one another unimpeded since, 
after all, they
represent free-particle wave functions. 
Describing this process in terms of effective forces would imply 
the presence of scattering and
accelerations, which do not occur here, and would be highly misleading.

Nonetheless, the concept of effective fermion repulsion is
evident in many texts, particularly in discussions of the 
behavior of an ideal fermion gas --- a case we explore further in section II.
A common usage of the repulsion
idea is in the interpretation of the second virial coefficient of an ideal
gas. 
The first correction to classical ideal gas pressure due
just to statistics is positive for spinless fermions and negative for
spinless bosons. 
Heer\cite{Heer} (similar to most other texts that treat the
subject, including one authored by one of us\cite{Brehm}) says, ``The
quantum correction that is introduced by statistics appears as an attractive
potential for BE [Bose-Einstein] statistics and as a repulsive potential for
FD [Fermi-Dirac] statistics.'' 
Patria's book\cite{Patria} carries the idea further, developing an actual
mathematical formula for the effective interaction between fermions or
between bosons! He says, ``In the Bose case, the potential is throughout
attractive, thus giving rise to a `statistical attraction' among bosons; in
the Fermi case, it is throughout repulsive, giving a `statistical repulsion'
among fermions.'' This 
 interaction formula first appeared in 1932 in
an article by Unlenbeck and Gropper,\cite{Uhlen} who may well be the
originators of the whole statistical interaction picture. We discuss this
formula in more detail in the next section.

Wannier\cite{Wannier} is a bit stronger
in his assessment of the quantum thermal distribution function for fermions:
``The particles exert a very strong influence on each other because a
particle occupying a state excludes the others from it. This is equivalent
to a strong repulsive force comparable to the strongest forces occurring in
the problem.''

The modern physics book of Leighton\cite{Leighton} omits the word
``effective'' in discussing the so-called fermion interaction: ``As compared
with the behavior of hypothetical but distinguishable particles, Bose
particles exhibit an additional attraction for one another and tend to be
found near one another in space; Fermi particles, on the contrary, repel one
another and tend not to be found near one another in space.'' The classic
kinetic-theory handbook by Herschfelder et al\cite{Hirsch} may set the
record by using the terms ``apparent repulsion'' (and ``apparent
attraction'') four times.

Griffiths\cite{Griffiths} 
does an interesting calculation of the average distance between
two particles at positions $x_{1}$ and $x_{2}$ when one is in state $\psi
_{a}$ and the other in $\psi _{b},$ the two functions being orthogonal and
normalized. For distinguishable particles with wave function $\psi
_{a}(x_{1})\psi _{b}(x_{2}),$ the mean-square separation is
\begin{equation}
\langle (x_{1}-x_{2})^{2}\rangle _{d}=\langle x^{2}\rangle _{a}+\langle
x^{2}\rangle _{b}-2\langle x\rangle _{a}\langle x\rangle _{b},
\end{equation}
where $\langle x^{2}\rangle _{i}=\int dx\,x\,\left| \psi _{i}(x_{1})\right|
^{2}.$ For spinless fermions the wave function must be antisymmetrized, and
for bosons symmetrized, giving
\begin{equation}
\Psi =\frac{1}{\sqrt{2}}\left[ \psi _{a}(x_{1})\psi _{b}(x_{2})\pm \psi
_{a}(x_{2})\psi _{b}(x_{1})\right],   \label{SymWF}
\end{equation}
where the upper sign is for bosons and the lower for fermions.  From this
form it is easy to compute the corresponding mean-square separation as
\begin{equation}
\langle (x_{1}-x_{2})^{2}\rangle _{\pm }=\langle (x_{1}-x_{2})^{2}\rangle
_{d}\mp 2\left| \langle x^{2}\rangle _{ab}\right|, 
\end{equation}
where $\langle x\rangle _{ab}=\int dx\,x\,\psi _{a}^{*}(x)\psi _{b}(x)$.
Thus he finds that bosons tend to be closer together and fermions farther
apart when compared to distinguishable particles. Griffiths comments, ``The
system behaves as though there were a `force of attraction' between
identical bosons, pulling them closer together, and a `force of repulsion'
between identical fermions, pushing them apart. We call it an exchange
force, although it's not really a force at all---no physical agency is
pushing on the particles; rather it is a purely a geometrical consequence of
the symmetrization requirement.'' This wording shows more care than the
works cited above and is thus less likely to be
misinterpreted. However, the term ``force'' has explicit meaning for
physicists. It implies a push or pull, 
along with its associated acceleration,
deflection, scattering, etc. Are these elements properly associated with
the exchange force? If not, then the term should be replaced with something
carrying more accurate connotations.  

Our intention is not to be critical of authors for using the words
``repulsion'' and ``attraction'' in describing the statistical effects of
wavefunction antisymmetry or symmetry. This concept has been with physics
since the early days of quantum mechanics. Nevertheless, it is important
to examine the
usefulness of this heuristic interpretation of the mathematics.  
As Layzer has
pointed out, no such interpretation can carry the whole weight of the
rigorous mathematical formulation; however, if a heuristic interpretation
brings along the baggage of subsequent misconceptions, 
then physicists must be more circumspect in its
use. 

For
example, consider the following case where there is a complete breakdown of
the concept. Suppose two spinless fermions or bosons have a completely
repulsive interparticle potential and impinge on one another at energies low
enough that there is only s-wave scattering. As we
show in Sec.~\ref
{sec:BADIDEA}, if the scattering amplitude for
distinguishable particles is $f$, then the scattering amplitude for fermions
vanishes identically, whereas for that for bosons is $2f$. 
In this case statistical symmetry has \emph{diminished} the
interaction for fermions---not made it more repulsive--and it has \emph{%
enhanced} the interaction for bosons---not made it less repulsive.

Wherever the idea of an effective force breaks down (as
it does in our wave-packet description and in the s-wave scattering
example), we need to replace this interpretation with other
heuristic interpretations that better represent the physics. This is our
aim in each of the examples we analyze below.

In Sec.~\ref{sec:repulsion} we examine more closely the physics that gives
rise to the idea of an effective statistical interaction between quantum
particles and will derive the Uhlenbeck-Gropper formula for the interaction.
Section~\ref{sec:BADIDEA} will take the opposite point of view, and find
cases where the idea is highly misleading and indeed where the effect is
actually opposite the usual implication. 
Sec.~\ref{sec:DISCUSS} summarizes our conclusions.

\section{EXAMPLES OF THE STATISTICAL INTERACTION}

\label{sec:repulsion} There are several places where the idea of a
statistical interaction arises somewhat naturally,
and might seem to imply an effective force. 
The virial correction to
the pressure of an ideal gas is most likely the origin of this idea of
effective interaction. The physics of white dwarf stars is another classic
example of ``Fermi repulsion.'' The diatomic hydrogen atom is bound in the
electron singlet state, while the triplet is unbound, which is often used as
an example of the effective repulsion between like-spin electrons due to the
Paul principle. When two rare gas atoms approach one another there
is a hard-core exponential repulsion between the atoms, which is often
explained by the electron statistical repulsion. On the other hand when
trapped bosons condense, they collapse to a smaller region in the center of
the trap; which gives the impression of an effective boson statistical
attraction. In each of these cases we will show that relying on the
intuitive idea of Pauli repulsion or Bose attraction may actually hinder
understanding of the basic phenomena. Alternative explanations
are provided. \vspace{0.2in}

\emph{Virial expansion.} A real gas has an equation of state that differs
from that of an ideal classical gas. For high temperature $T$ and low
density $n$ of the gas, the pressure $P$ can be written
\begin{equation}
P=nk_{B}T\left( 1+nB(T)\right),  \label{Virial}
\end{equation}
where $k_{B}$ is the Boltzmann constant and $B$ is known as the second
virial coefficient. This equation gives the lowest terms of the
so-called virial expansion, a series in powers of $n\lambda ^{3},$ where $%
\lambda $ is the thermal wavelength, 
given by
$\lambda =\sqrt{h^{2}/(2\pi mk_{B}T)}$ for particles of mass $m$.

For ideal spinless fermions and bosons, standard calculations\cite
{Patria,Huang} give the effect of Fermi or Bose symmetry: 
\begin{equation}
B(T)=-\eta \frac{\lambda ^{3}}{2^{5/2}},  \label{FBsecvir}
\end{equation}
where $\eta =\pm 1,$ with the plus sign for bosons and the minus for
fermions. Thus fermions exert a larger pressure, and bosons a smaller
pressure, on the walls than a classical gas at the same temperature.

Compare this result with that for a classical interacting gas, where the
second virial coefficient is given by\cite{Patria} 
\begin{equation}
\label{Bcl}
B(T)=\frac{1}{2}\int d\mathbf{r}\left( 1-e^{-\beta U(\mathbf{r})}\right),
\end{equation}
where $U(r)$ is the real interatomic potential at separation $r$ and $\beta
=1/k_{B}T.$ It is evident from Eq.~(\ref{Bcl}) that a
completely repulsive potential leads to a positive $B(T)$ and a positive
contribution to the pressure, while an attractive one results in a negative
contribution.

A connection to the Fermi or Bose ideal gas is made by considering the pair
density matrix given by 
\begin{equation}
G(1,2)=V^{2}\frac{\langle \mathbf{r}_{1}\mathbf{r}_{2}|e^{-\beta H_{12}}|%
\mathbf{r}_{1}\mathbf{r}_{2}\rangle }{\mathrm{Tr}\left( e^{-\beta
H_{12}}\right) }=\lambda ^{3}\sum_{\mathbf{p}_{1}\mathbf{p}_{2}}\psi _{%
\mathbf{p}_{1}}(\mathbf{r}_{1})\psi _{\mathbf{p}_{2}}(\mathbf{r}%
_{2})e^{-\beta (\epsilon _{p_{1}}+\epsilon _{p_{2}})}(1+\eta P_{12})\psi _{%
\mathbf{p}_{1}}^{*}(\mathbf{r}_{1})\psi _{\mathbf{p}_{2}}^{*}(\mathbf{r}_{2}),
\label{W}
\end{equation}
where $\psi _{\mathbf{p}_{i}}(\mathbf{r}_{i})$ is a plane-wave momentum
state for particle $i$ and $P_{12}$ is the permutation operator
interchanging $\mathbf{r}_{1}$ and $\mathbf{r}_{2}$. The single-particle
energy is $\epsilon _{p}=p^{2}/2m.$ Changing the momentum sums to integrals
and carrying out the calculations leads us to the following result depending
on relative position $r_{12}$ only: 
\begin{equation}
G(r_{12})=\left( 1+\eta e^{-2\pi r_{12}^{2}/\lambda ^{2}}\right) .
\label{TwoBdydens}
\end{equation}
The purely classical ideal gas result would correspond to $\eta =0$ 
with no correlation between particles.  Fermions, on the other hand, 
have $G$ small within a thermal wavelength, an example of the spatial 
consequences of the Pauli principle.  Bosons have $G$ larger 
than the classical value. This 
result is consistent with the Griffiths' calculation of $\langle 
(x_1-x_2)^{2}\rangle $ cited in Sec.~\ref{sec:Intro}.

Spatial correlations in a classical gas are described by the two-particle
distribution function given by $G_{cl}(1,2) = e^{-\beta U(\mathbf{r})}$. Thus, a
s 
in Ref.~%
\onlinecite{Uhlen} and repeated in Ref.~\onlinecite{Patria}, we identify by
analogy an effective statistical potential as
\begin{equation}
U_{\mathrm{eff}}(r)=-k_{B}T\ln \left( 1+\eta e^{-2\pi r_{12}^{2}/\lambda
^{2}}\right) .  \label{UGpot}
\end{equation}
This quantity is plotted in Fig.~\ref{fig:UGpot}. 
It is purely repulsive for fermions and
attractive for bosons. If we put this back into the classical expression for
the second virial coefficient, we get precisely the result quoted in Eq.~(%
\ref{FBsecvir}). A repulsive potential excludes atoms from approaching too
closely and raises the pressure; fermions also have an ``excluded volume''
of $\lambda ^{3}$ 
and an increase in pressure.
This comparison seems to be the major impetus behind the 
concept of ``effective force'' as applied to
Fermi statistics.
Is the physics similar enough for the analogy to be
useful? Our opinion is that it is not very helpful, as we argue below.

In a classical gas the rms average momentum remains $\sqrt{%
\overline{p^{2}}}=$ $\sqrt{3mk_{B}T}$ even when there are interactions.
Pressure is force per unit area and the force comes from the impulse of an
atom striking the wall. The average force that a single particle exerts on
the wall is, by the impulse-momentum theorem, $F=\Delta p/\Delta t$ where $%
\Delta p$ is twice the average momentum and $\Delta t$ is the average time
over which the force is exerted, which here is \emph{not} the time of
contact, but rather the time for an atom to cross the width $L$ of the
container, that is, $\Delta t=mL/\overline{p}$. When one makes the volume of
an ideal classical gas smaller (at constant $T)$ $\overline{p}$ is
unchanged, but the transit time $\Delta t$ is diminished causing the
pressure to increase$.$ Analogously, when a classical gas has repulsive
interactions ``turned on'' with no change in the temperature or $\overline{p}
$, the pressure rises because of a decreased average transit time: some
molecules bounce off others back to the wall they just left. But this is 
\emph{not} what happens in the fermion case. The idea that the correlation
hole in the two-body density Eq.~(\ref{TwoBdydens}) gives rise to
``bounces'' or deflections of fermions from one another is a misconception
that arises from the idea of a Pauli repulsion. 
When one compares Fermi gas dynamics to that of classical statistics,
what is altered is \emph{not} the effective $L$ in the transit time, but
rather the $\overline{p}$ in both $\Delta p$ and in $\Delta t.$ For a given
value of $T$, the momentum distribution in an ideal Bose or Fermi
gas differs 
from that in an ideal classical gas. The exact quantum second virial
coefficient is given by\cite{Huang} 
\begin{equation}
B(T)=\frac{1}{2V}\int d\mathbf{r}_{1}d\mathbf{r}_{2}\left[ 1-G(1,2)\right] .
\label{QuantB}
\end{equation}
This result explains why the substitution of $U_{\mathrm{eff}}$ into the
classical equation gives the exact answer. Nevertheless, it is not the
spatial dependence of $G$ that gives us physical insight; it is the \emph{%
momentum} dependence: Carry out the position integration indicated in Eq.~(%
\ref{QuantB}) with $G$ as given by Eq.~(\ref{W}). The result is 
\begin{equation}
\frac{1}{2V}\int d\mathbf{r}_{1}d\mathbf{r}_{2}G(1,2)=\frac{\lambda ^{6}}{V}%
\left\{ \frac{1}{2}\left[ \sum_{\mathbf{p}_{1}\mathbf{p}_{2}}e^{-\beta
(\epsilon _{p_{1}}+\epsilon _{p_{2}})}+\eta \sum_{\mathbf{p}}e^{-2\beta
\epsilon _{p}}\right] \right\} .
\end{equation}
The quantity inside the curly brackets is the partition function for 
two quantum particles.  The first term of this is just the classical 
partition function and its contribution is already accounted for in 
the classical ideal gas pressure; it cancels out in 
Eq.~(\ref{QuantB}).  The second term corrects the wrong classical 
momentum distribution represented by the first term.  The classical 
term includes double-occupation states; for fermions the second term 
cancels these out.  For bosons, the classical counting undercounts 
these double-occupation terms and the second term corrects that fault 
as well.  Writing the second virial coefficient in momentum space 
clarifies how the change in momentum distribution affects the 
pressure.  For bosons, there is a lowering of the average momentum so 
the force on the wall is lessened.  For fermions, the momentum is 
raised increasing the pressure.  The idea of an effective repulsion 
between fermions just ignores the real physics and gives a very 
poor analogy with classical 
repulsive gases.  \vspace{0.2in}

\emph{White dwarf stars and related objects.} It is the fermion 
zero-point pressure that prevents the collapse under gravitational 
forces of the white dwarf star.  Krane\cite{Krane} says, ``A white 
dwarf star is prevented from collapse by the Pauli principle, which 
prevents the electron wave functions from being squeezed too close 
together.\ldots Will the repulsion of the electron wave functions due 
to the Pauli principle be able to prevent the collapse of any star, no 
matter how massive?'' (This line leads into a discussion of neutron 
stars.) We feel this qualitative picture of what goes on in a white 
dwarf star could, as with the second virial coefficient 
interpretation, be greatly improved by discussion in terms of the 
momentum-space features of the Pauli principle.  Most elementary 
discussions\cite{Krane} of white dwarfs incorporate a discussion of 
``Fermi repulsion'' by doing a dimensional analysis that equates the 
zero-point energy of the ideal Fermi gas to the gravitational 
self-energy of the star matter.  The Fermi temperature is much greater 
than the physical temperature in the star so that the $T=0$ fermion 
gas is used as a model.

An alternate physical description arises from considering the hydrostatic
equilibrium conditions of the star.\cite{Chandra} The star is assumed to
contain $N$ nuclei (assumed all helium) in radius $R$. 
A spherical shell of thickness $dr$ at radius $r$  
has an outward force due to a difference
between the pressure $P(r)$ on the inner surface and the pressure $%
P(r+dr)=P+dP$ (with $dP<0)$ on the outer surface, caused by the
nonuniform nuclear number density of the star, $n(r)$. 
This net outward force $4\pi
r^{2}dP$ is balanced by the gravitational pull toward the center due to the
total mass $M(r)$ enclosed by the shell. The mass of the shell itself is $%
4\pi r^{2}n(r)\,dr\,m_{\mathrm{He}},$ where $m_{\mathrm{He}}$ is the helium
mass, so that 
\begin{equation}
dP=-\frac{GM(r)n(r)\,dr\,m_{\mathrm{He}}}{r^{2}}.  \label{diffe}
\end{equation}
The crucial idea is that $P$ is the pressure of a degenerate electron gas
with the electron density maintained by charge neutrality at twice the
helium number density $n_{\mathrm{e}}(r)=2n(r).$ For a non-relativistic
model the Pauli pressure at $T=0$ is given by standard statistical arguments%
\cite{Huang} as $P\approx \hbar ^{2}n_{\mathrm{e}}^{5/3}/m_{\mathrm{e}}.$
Chandrasekhar\cite{Chandra} develops a second-order differential equation
for $n(r)$ from these steps. We can do a simple dimensional analysis based
on Eq.~(\ref{diffe}) by replacing $dP/dr$ by $-P/R,$ $n(r)$ by $N/R^{3},$ $%
M(r)$ by $M(R)=m_{\mathrm{He}}N$, etc.\ to arrive at 
\begin{equation}
R=\frac{\hbar ^{2}}{Gm_{\mathrm{e}}m_{\mathrm{He}}^{2}}\frac{1}{N^{1/3}}%
\approx \frac{1}{M^{1/3}}.
\end{equation}
This is the usual non-relativistic result, which does not demonstrate the
collapse at some large $M$ like the relativistic case, but gives the idea
behind the stability of the star.

The gravitational attraction on a mass element is balanced by the difference in
Pauli pressure across the mass shell. 
In order to develop a qualitative argument
for the strong density dependence of the Pauli pressure that
supports the star against gravitational collapse,
we can return to the argument used for the virial coefficient. 
In a box of
sides $L$, the pressure is force per unit area $A$, or $P=(N/A)\Delta p/%
\Delta t.$ But the average momentum per particle $\Delta p$
imparted to the wall for a
degenerate Fermi gas is of order $p_{\mathrm{F}},$ the Fermi momentum. The
transit time is $\Delta t$ $\sim Lm_{\mathrm{e}}/p_{\mathrm{F}}$ so that 
\begin{equation}
P\approx \frac{N}{AL}\frac{p_{\mathrm{F}}}{m_{\mathrm{e}}/p_{\mathrm{F}}}=n_{%
\mathrm{e}}\frac{p_{\mathrm{F}}^{2}}{m_{\mathrm{e}}}.
\end{equation}
The Fermi momentum itself is strongly dependent on the density because of
the necessity to fill the single-particle energy levels with two per momenum
state. This requirement is $N=(2V/h^{3})\int d\mathbf{p\,}n_{\mathbf{p}}$
with $n_{\mathbf{p}}$ a step function cutting off at $p=p_{\mathrm{F}}.$
This integral gives $p_{\mathrm{F}}=\hbar (3\pi ^{2}n_{\mathrm{e}})^{1/3}.$
Note that $p_{\mathrm{F}}$ is related to a deBroglie wavelength by 
\begin{equation}
p_{\mathrm{F}}=\frac{\hbar }{\lambda }\approx \hbar n_{\mathrm{e}}^{1/3}.
\end{equation}
Thus the maximum wavelength is approximately the interparticle separation,
which one can argue is necessitated by the Pauli principle requiring that
the electrons be in single-particle wave packets compact enough that 
they don't overlap. This is an argument about quantum-mechanical
wave-function correlation rather than an argument based on an effective
force. The connection to the Pauli pressure is the high momentum that this
correlation induces. We end up with 
\begin{equation}
P\,\approx \,n_{\mathrm{e}}\frac{p_{\mathrm{F}}^{2}}{m_{\mathrm{e}}}\,%
\approx \,\frac{\hbar ^{2}}{m_{\mathrm{e}}}n_{\mathrm{e}}^{5/3}.
\end{equation}

If by ``preventing the wave functions from being squeezed too close 
together''\cite{Krane} one means that the fermion wave function must 
have sufficient curvature for nodes to appear whenever any two 
coordinates are equal, then the idea leads directly to the correct 
behavior.  This extra curvature requires higher Fourier components.  
The pressure differs from one kind of statistics to another directly 
because of differing momentum distributions; the Fermi distribution 
involves larger average momenta, giving it a Pauli pressure.  The idea 
of ``wave function repulsion'' as a \emph{correlation} that leads to 
this momentum distribution might be useful, although the word 
``repulsion'' still carries the connotation of a force, which is less 
useful.

The physical explanations of neutron stars,\cite{stars} strange quark
matter,\cite{Witten} 
the Thomas-Fermi model of the atom,\cite{Harris} are all
analogous to the white dwarf star in that the Pauli pressure 
of a Fermi fluid is the basis of resistance to compression.
\vspace{0.2in}

\emph{The hydrogen molecule and interatomic forces}. The singlet electron
state of hydrogen is bound while the triplet state is unbound. Is it a case of
the Pauli repulsion giving the spatially antisymmetric state associated with
the triplet higher energy? Griffiths,\cite{Griffiths} applying the
discussion of exchange forces to this problem, says ``The
system behaves as though there were a `force of attraction' between
identical bosons, pulling them closer together.\ldots If electrons were
bosons, the symmetrization requirement\ldots would tend to concentrate the
electrons toward the middle, between the two protons\ldots , and the
resulting accumulation of negative charge would attract the protons inward,
accounting for the covalent bond.\ldots But wait. We have been ignoring
spin.\ldots '' He then talks of the fact that the entire spin and space wave
function must be antisymmetric and gets the proper bonding in the singlet
state. He shows that for the spatially antisymmetric triplet state ``the
concentration of negative charge should actually be shifted to the
wings\ldots, tearing the molecule apart.''

While this explanation is very carefully worded 
and provides a very useful physical picture of the hydrogen
bond, a strikingly different picture of covalent bonding and antibonding is
given by the work of Herring.\cite{Herring} Herring argues that the energy
difference between singlet and triplet states (in widely separated atoms at
least) is properly interpreted as a splitting between atomic levels due to
tunneling. 
Consider the hypothetical case of two
\emph{spinless, distinguishable} electrons in a hydrogen molecule.
The Hamiltonian has the form 
\begin{equation}
H=t_{1}+t_{2}+V(12)+U(1)+U(2),
\end{equation}
in which $t_{i}$ is the kinetic energy operator for particle $i,$ $V(12)$
represents the particle-particle interaction, and $U(i)$ is 
an external double-well potential 
representing the attraction of the $i^{th}$ electron to the two nuclei located, 
say, at $R_{a}$ and $R_{b}$.
The Hamiltonian is symmetric under interchange of the two
particles, so the eigenfunctions must be either symmetric or antisymmetric,
even for these distinguishable particles. Let $\psi _{+}$ and $\psi _{-}$
represent the lowest symmetric and antisymmetric eigenfunctions,
respectively, with corresponding energies $E_{+}$ and $E_{-}.$

The combination 
\begin{equation}
\phi _{ab}(1,2)=\frac{1}{\sqrt{2}}\left( \psi _{+}+\psi _{-}\right) 
\end{equation}
is a function for which particle 1 is localized near site $R_{a,},$ and
particle 2 near site $R_{b.}$ If $P_{12}$ is the permutation
operator then the function $P_{12}$ $\phi _{ab}(1,2)=\phi _{ab}(2,1)=\phi
_{ba}(1,2)$ is localized around the exchanged sites; that is, particle 1 is
localized near site $R_{b,}$ and particle 2 near site $R_{a}.$ Herring
calls the functions $\phi _{ab}(1,2)$ and $\phi _{ba}(1,2)$ ``home-base
functions.'' If one sets the initial conditions such that the particles are
in $\phi _{ab}(1,2),$ then the two particles will tunnel through the
double-well barrier between $\phi _{ab}(1,2)$ and $\phi _{ba}(1,2)$ with
frequency $\omega =\left( E_{+}-E_{-}\right) /\hbar .$ We can write the
energy of these two lowest states for distinguishable particles as 
\begin{equation}
E_{\pm }=E_{0}\pm J ,  \label{Edist}
\end{equation}
where $E_{0}=(E_{+}+E_{-})/2$ and $J=(E_{+}-E_{-})/2.$ 
A theorem\cite{Lieb} states that the symmetric nodeless 
state must be the ground state, thereby implying that $J$ is negative.
This energy splitting occurs independent of any spin effects.

If the two particles are spin-1/2 fermions, exactly the same physics holds,
except now the symmetric wave function $\psi _{+}$ must be associated with
an antisymmetric spin function while $\psi _{-}$ must be associated with a
symmetric spin function in order to keep the entire wave function overall
antisymmetric. The result is that Eq.~(\ref{Edist}) is replaced by 
\begin{equation}
E_{\pm }=E_{0}\pm J\mathbf{\sigma }_{1}\cdot \mathbf{\sigma }_{2} ,
\label{Eferm}
\end{equation}
where $\mathbf{\sigma }_{i}$ is a Pauli spin matrix.  This operator 
expression acts in spin space to associate the correct spin state 
(singlet or triplet) with the correct energy.\cite{spinperm} Since the 
symmetric spatial state is the ground state, as in the distinguishable 
particle case, the singlet state energy of the electrons is lower than 
that of the triplet state.

Note that we are not at all criticizing the idea quoted from Ref.~%
\onlinecite{Griffiths} that the lowering of the energy in the spin singlet
state can be associated with the concentration of the electron cloud in the
region between the two nuclei. However, the energy lowering would arise even if
the two particles were distinguishable; so it does not actually stem from
their fermion character.\cite{surface} Of course, the fact that the
corresponding spin state must be singlet \emph{is} a fermion property. The
suggestion that Fermi statistics or Pauli repulsion plays a role in the
lowering of the singlet relative to the triplet state of H$_{2}$ misses the
essential fact that much of the energy difference is due to splitting
between tunneling states and that the tunneling ground state must be
nodeless and symmetric.

Let's continue the above discussion, but with the hydrogen nuclei replaced
with helium nuclei. We can get an idea of the behaviour of the electronic
energy for this pair of helium atoms by using the same symmetric and
antisymmetric wave functions. Because of the Pauli principle, the two
extra electrons would (in some approximation) be placed in the
spatially antisymmetric, antibonding, triplet state, thereby losing the
tunneling energy advantage of the symmetric state. This extra energy
supplies a physical explanation for the repulsive interatomic interaction
when the closed-shell electron clouds start to overlap. Within the
Born-Oppenheimer approximation\cite{Hirsch,Ziman} the electronic energy
(plus the internuclear Coulomb repulsion) is used as a potential energy for
the atomic \emph{nuclei}. The short-ranged repulsive part of this
interaction potential between two rare-gas atoms is often described by a
phenomenological $1/r^{12}$ or exponential repulsion.\cite{Hirsch,dispersion}
Here then is a case of a real repulsion arising; however, it is a repulsion
between the nuclei---not the electrons.

While the Pauli principle is certainly vital in understanding molecular
forces, the idea of an effective fermion statistical repulsion has never
really entered the picture. Indeed, we feel its introduction short-circuits
the discussion and could cause one to miss the basic physics. \vspace{0.2in}

\emph{Bose-Einstein condensation}. Condensation of bosons into a harmonic
trap might seem the best example of boson effective attraction.\cite{Pethick}
The condensate in a trap is a noticably smaller object than the cloud of
non-condensed atoms surrounding it. Of course, the real reason for this is
that the ground state in the trap of the interacting bosons has smaller
radius than that of the excited states. The particles are correlated to be
in the same state; in this case it is a spatially more compact one.

One of the present authors has also used the following argument\cite{Brehm}
to explain the fact that the lowest excited mode in a Bose fluid is a phonon
rather than a single-particle motion: ``Bosons prefer to be in the same
state with one another, so that if one atom is pushed on by an external
force, all the particles within a deBroglie wavelength $\lambda $ (which is
large at low temperature) want to move in the same way. The collective
motion of a sound wave allows this while the single-particle motions are
frozen out by this tendancy.'' This argument seems based on the idea of a
kind of boson effective attraction. A more rigorous argument is given by
Feynman.\cite{Feynman} If $\Phi $ is the ground state of the Bose fluid,
then one might suppose that $e^{i\mathbf{k}\cdot \mathbf{r}_{1}}\Phi $ is an
excited state involving a single particle with momentum $\mathbf{k}.$
However the state has to be symmetric, so this state must be replaced by $%
\sum_{i}e^{i\mathbf{k}\cdot \mathbf{r}_{i}}\Phi $, which is precisely the
one-phonon state. The particles ``preferring to be in the same state'' is a
verbal expression to represent wave-function symmetry. Superfluids can be
described by a ``wave function'' (order parameter), which depends on a
single position variable, has a magnitude and phase, and represents the
superfluid distribution. It costs energy to make this function nonuniform,
as when a vortex is present. The system ``prefers'' to have the same phase
and amplitude throughout, a property sometimes called ``coherence.'' Any
idea of an effective boson attraction is better replaced by this latter
concept.

\section{WHERE THE IDEA OF A STATISTICAL INTERACTION FAILS}

\label{sec:BADIDEA} We have already argued above that the idea of a
statistical fermion repulsion or boson attraction has the potential to make
one miss the essential physics of the physical effect being explained. Worse
however, is the fact that this 
idea might cause misconceptions
and lead to incorrect 
conclusions. We present here some cases where that might occur.
\vspace{0.2in}

\emph{The other spin state}.  Most of the textbooks quoted in 
Sec.~\ref{sec:Intro} say unequivocally that fermions repel and bosons 
attract without the qualification of, say, the 
term ``spinless.'' These books have ignored, at some 
pedagogic risk, the effects of spin, which is usually taken into 
account only later.  
The effective repulsion or attraction (if there 
were one) is an effect of the spatial part of the wave function only.
If the total spin state is symmetric, the space wave function is antisymmetric
for fermions and symmetric for bosons, leading to the 
effects envisioned in most textbooks. 
However, when the total spin state is antisymmetric
(as for two spin 1/2 particles in a spin singlet, or for two spin 1 particles
in the $S=1,m_{s}=0$ state) the roles of fermions and bosons are
reversed. Two spin 1/2 fermions in the spin singlet state behave like two
spinless bosons, and two spin 1 bosons in the $S=1,m_{s}=0$ state behave
like spinless fermions.
\vspace{0.2in}

\begin{figure}[tbp]
\centering
\includegraphics[width=5.5in]{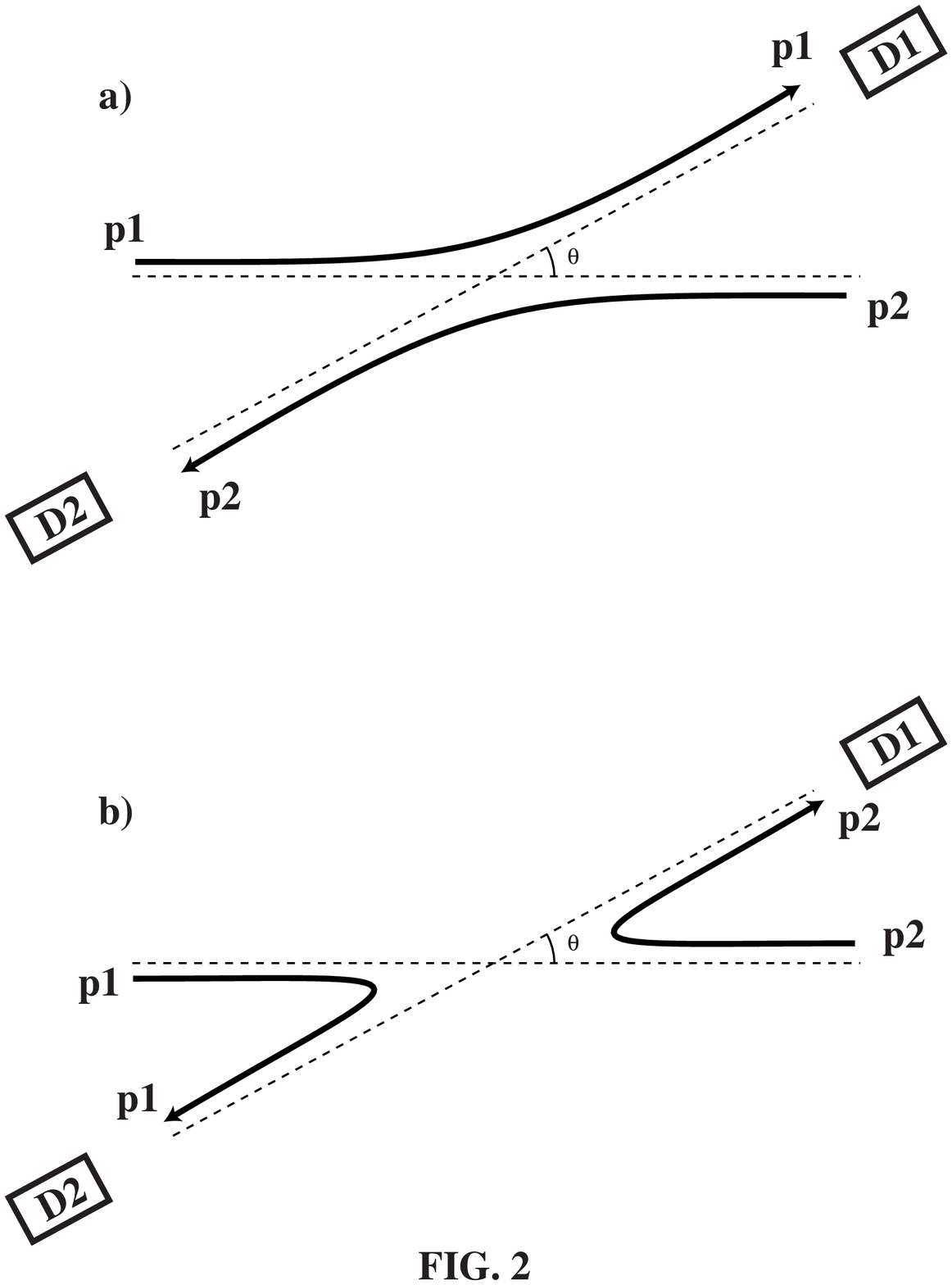}
\caption{Diagrams for two-particle scattering. }
\label{fig:twopartscatt}
\end{figure}

\emph{Scattering theory}. When two particles scatter elastically via a
repulsive force the idea of an additional effective interaction due to Fermi
or Bose symmetry can lead to trouble. In the center-of-mass frame the two
particles approach from opposite directions and scatter into opposite
directions as shown in Fig. \ref{fig:twopartscatt}a. If the particles are
distinguishable, the probability of detecting particle p1 in detector D1 and
particle p2 in detector D2 is given by
\begin{equation}
P(\mathrm{p1\ in\ D1})=|f(\theta )|^{2},
\end{equation}
where $f(\theta )$ is the scattering amplitude\cite{philess}. Similarly, the
probability of detecting particle p2 in detector D1 and particle p1 in D2
(as in Figure \ref{fig:twopartscatt}b) is given by
\begin{equation}
P(\mathrm{p1\ in\ D2})=|f(\pi -\theta )|^{2}.
\end{equation}
When you don't care which particle goes to which detector, but just want to
measure the cross section for either particle in a detector, the probability
for a particle in detector D1 is
\begin{equation}
P(\mathrm{p1\ or\ in\ D1})=|f(\theta )|^{2}+|f(\pi -\theta )|^{2}.
\end{equation}
Since the particles are in principle distinguishable there is no
interference between amplitudes, even if the detectors themselves do not
identify the difference between particles.

Now suppose the two particles are indistinguishable. In this case the two
amplitudes corresponding to Figs.~\ref{fig:twopartscatt}a and \ref
{fig:twopartscatt}b interfere, and must be combined before squaring. If the
particles are identical fermions, the two-particle wave function is
antisymmetric with respect to particle exchange. Since diagrams \ref
{fig:twopartscatt}a and \ref{fig:twopartscatt}b are related by the exchange
of the two particles in the final state, they contribute to the total
amplitude with opposite signs. Thus, the probability to detect a fermion in
detector D1 is
\begin{equation}
P_{\mathrm{Fermi}}(\mathrm{p\ in\ D1})=|f(\theta )-f(\pi -\theta )|^{2}.
\end{equation}
This is obviously different from the distinguishable case above. The
difference is especially remarkable at $\theta =\pi /2$, where the fermion
scattering probability vanishes. Moreover, in the limit of s-wave
scattering, which is a good approximation for some low energy
cases, the scattering is independent of the angle $\theta $ and the fermion
probability for scattering is zero for all angles. A similar argument holds
for bosons, but with the amplitudes adding instead of subtracting, leading
to the scattering probability: 
\begin{equation}
P_{\mathrm{Bose}}(\mathrm{p\ in\ D1})=|f(\theta )+f(\pi -\theta )|^{2}.
\end{equation}
In this case, the scattering probability at $\theta =\pi /2$ is twice the
value for distinguishable particles. For s-wave scattering $P_{\mathrm{Bose}}
$ is a factor of two times the distinguishable value at all angles.

The interpretation of these results in the context of an effective fermion
repulsion or an effective boson attraction is quite confusing. For
scattering at 90 degrees, or for s-wave scattering at all angles,
it looks as if the total repulsive force is
reduced in the case of fermions (leading to a smaller scattering
probability) and enhanced in the case of bosons (leading to a larger
scattering probability). This is exactly backwards from the notion that the
scattering force should be supplemented by an effective repulsion for
fermions and partially canceled by an effective attraction for bosons. It
clearly demonstrates why the idea of an effective repulsion or attraction is
a dangerous concept.

Focusing on the direct effects of the Bose or Fermi symmetry leads to a more
useful conceptual approach to scattering. For two identical particles, the
total spin state is symmetric. For fermions having a total spin state that
is symmetric (either both spins up or both spins down), the space wave
function itself must be antisymmetric as written in Eq. (\ref{SymWF}). For
this wave function, the amplitude for the two fermions to be in the same
place ($r_{1}=r_{2}$) is obviously zero. As noted in 
Sec.~\ref{sec:Intro}, two
identical fermions are on average farther apart than two distinguishable
particles would be under the same circumstances. Consequently, the fermions
interact \emph{less} and are less likely to scatter. One can similarly argue
that bosons are closer together on average, interact more and are \emph{more}
likely to scatter.

This conclusion is true whether the scattering force is repulsive or
attractive, but it depends critically on the spin state of the two
particles. For identical particles the spin state is necessarily symmetric,
forcing the fermion spatial wave function to be antisymmetric or the boson
wave function to be symmetric. However, if the two particles are in an
antisymmetric spin state, e.g., two fermions in a spin-zero state, the
conditions are reversed.

As a specific example of repulsive scattering, 
consider the quantum electrodynamic interaction of two
electrons (Moller scattering). In QED, there are two lowest-order Feynman
diagrams that contribute to the scattering
amplitude with opposite signs, corresponding to the direct and exchange
diagrams of Figure \ref{fig:twopartscatt}. In the non-relativistic limit,
the electron spins do not change as a result of the interaction. (This is
due to the fact that the low energy interaction occurs primarily via an
electric field.) There is then only one final spin state to consider when
doing the calculation. If the initial state has both particles with spin up
then the final state also has both spins up. This case is treated in
introductory particle-physics texts\cite{HalzenMartin} and the cross section
for scattering is 
\begin{equation}
{\frac{d\sigma }{d\theta }}(\mathrm{identical\ spins})={\frac{m^{2}\alpha
^{2}}{32p^{4}}}\left( {\frac{1}{\sin ^{2}{\frac{\theta }{2}}}}-{\frac{1}{%
\cos ^{2}{\frac{\theta }{2}}}}\right) ^{2},
\end{equation}
where $\alpha $ is the fine structure constant and $p$ is momentum of the
electrons in the center-of-mass frame. One notes that the cross section at $%
\theta =\pi /2$ vanishes, as it should.

To explore the case of an antisymmetric spin wave function, one can also
calculate the scattering cross section for electrons in a spin-zero state:
\begin{equation}
{\frac{d\sigma }{d\theta }}(\mathrm{spin\ zero})={\frac{m^{2}\alpha ^{2}}{%
32p^{4}}}\left( {\frac{1}{\sin ^{2}{\frac{\theta }{2}}}}+{\frac{1}{\cos ^{2}{%
\frac{\theta }{2}}}}\right) ^{2}.
\end{equation}
This differs from the identical spin case in the relative sign of the two
terms.

These results should be compared to what the cross section would be if the
two electrons were distinguishable. In that case, only the direct diagram of
contributes, and the cross section for
scattering with both spins up turns out to be the same as the spin-averaged
cross section for electron-muon scattering found in many texts, with the
muon mass set equal to the electron mass. After symmetrizing around $%
\theta=\pi/2$ to account for detectors that are sensitive to either
particle, the cross section can be written as 
\begin{equation}
{\frac{d\sigma}{d\theta}}(\mathrm{distinguishable\ electrons}) = {\frac{%
m^2\alpha^2}{32p ^4}} \left({\frac{1}{\sin^4{\frac{\theta}{2}}}} + {\frac{1}{%
\cos^4{\frac{\theta}{2}}}}\right) .
\end{equation}

\begin{figure}[tbp]
\centering
\includegraphics[width=4.5in]{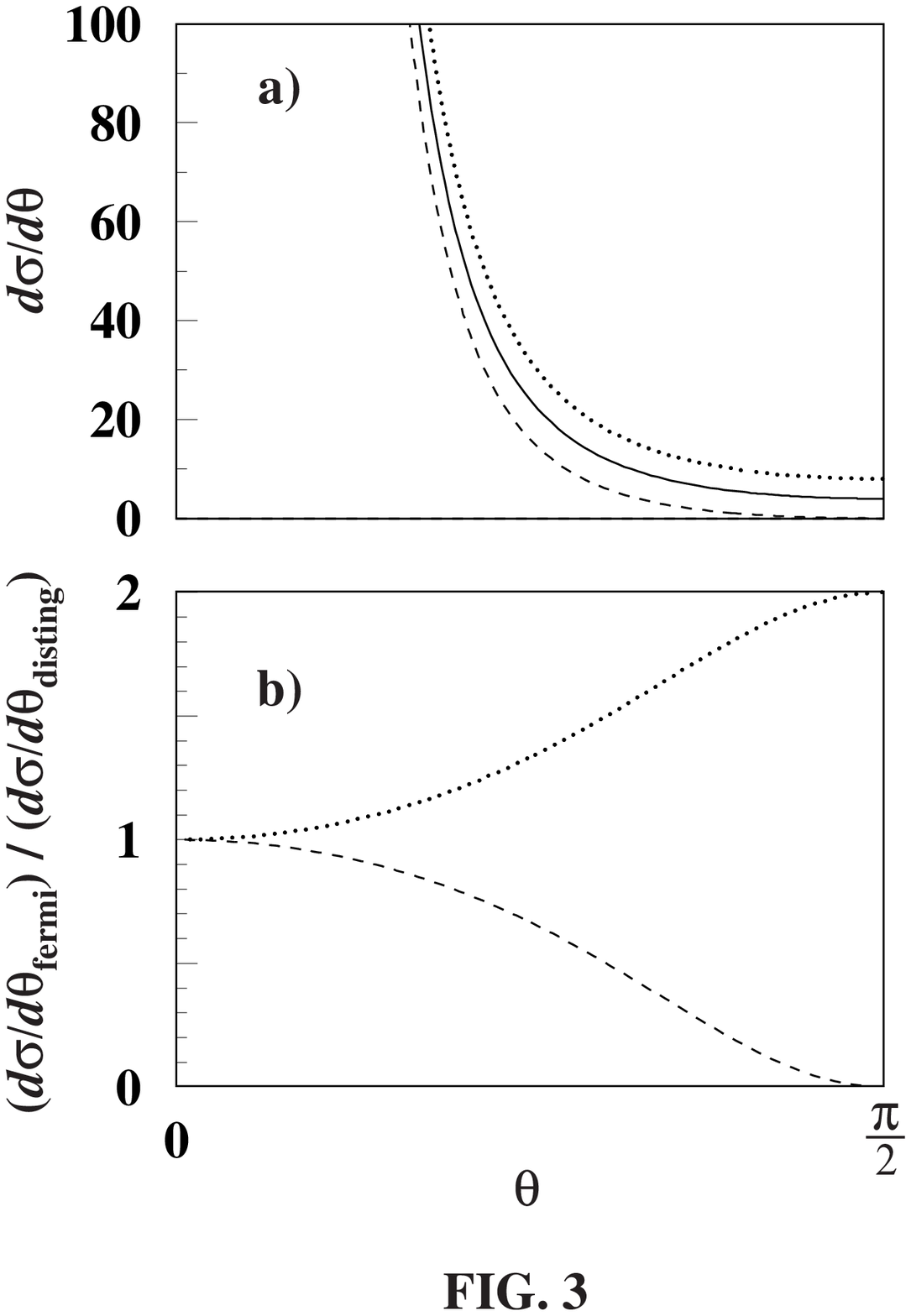}
\caption{
a) The cross section for electron scattering as a function of
scattering angle for identical spins (dotted line), for the spin-zero state
(dashed line), and for the hypothetical case of distinguishable particles
(solid line). All plots are in units of ${\frac{m^{2}\alpha ^{2}}{%
32p^{4}}}$.
b) The ratio ${\frac{d\sigma }{d\theta }}(\mathrm{identical})/{\frac{%
d\sigma }{d\theta }}(\mathrm{distinguishable})$ (dotted line) and the ratio
${\frac{d\sigma }{d\theta }}(\mathrm{spin\ zero})/{\frac{d\sigma }{%
d\theta }}(\mathrm{distinguishable})$ (dashed line). }
\label{fig:MollerXsection}
\end{figure}

All three cases are plotted as a function of scattering angle in Fig.~\ref
{fig:MollerXsection}a. All cross sections are symmetric around $\pi /2$, so
only the range 0 to $\pi /2$ is plotted. As expected, the symmetric spin
case gives the smallest scattering cross section, the antisymmetric spin
case gives the largest cross section, and the case of distinguishable
particles is in between. Moreover, as Fig.~\ref{fig:MollerXsection}b shows,
the ratios of the cross sections change as a function of scattering angle.
At small scattering angles, the fermion cross sections are almost the same
as the distinguishable particle cross section. The maximum difference occurs
at $\theta =\pi /2$. It is difficult for any kind of effective fermion
interaction to capture this effect, and moreover, the idea of an effective
Fermi repulsion gives the wrong sign in the case of a repulsive scattering
force.
\vspace{0.2in}

\emph{Transport theory}.  Consider the example of thermal conductivity 
in a polarized fermion gas.  One might think from the idea of Pauli 
repulsion that increasing polarization would \emph{shorten} the 
particle's mean-free path in the gas, which in turn would lower the 
thermal conductivity $\kappa $.  The opposite behavior is more likely 
to happen.  At sufficiently low temperature where s-wave scattering 
predominates, polarization will actually cause a dramatic increase in 
$\kappa $, because, as we have just seen, the s-wave scattering 
cross section between like-spin fermions vanishes and only scattering 
between unlike spins, which now happens less often, can contribute to 
the mean free path.

We treat a gas obeying Boltzmann statistics, but having full 
quantum-mechanical collisions.  For this to be true, the deBroglie 
wavelength must be larger than the scattering length, but smaller than 
the average separation between particles.  If the temperature is low 
enough s-wave scattering will predominate.  This situation can occur, 
for example, in trapped Fermi or Bose gases.  The heat current for 
spin species $\mu$ in temperature gradient $dT/dz$ is given by 
arguments analogous to those for an unpolarized gas\cite{Reif}:
\begin{equation}
\label{spincurrent}
J_{\mu }=-n_{\mu }\overline{v}l_{\mu }k_{B}\frac{dT}{dz},
\end{equation}
where $n_{\mu }$ is the density of $\mu $ spins, $\overline{v}$ is the
average velocity of either spin species, $l_{\mu }$ is the mean free
path of a $\mu $ spin and $k_{B}$, the Boltzmann constant, is the
specific heat per molecule.  In Eq.~(\ref{spincurrent}) $\mu$ is $+$
for up spins and $-$ for down spins; there is a separate heat equation
for each spin species.  We have dropped any constant factors in the
expression.  When s-wave scattering dominates, up spins can interact
only with down spins and
not with each other, and vice versa. Thus the mean free path is $l_{\mu }=%
\overline{v}\tau _{\mu }$ where $\tau _{\mu }$ is the inverse of the
scattering rate given by 
\begin{equation}
\frac{1}{\tau _{\mu }}=n_{-\mu }\overline{v}\sigma _{+-} ,  \label{rate}
\end{equation}
with $\sigma _{+-}$ the cross section for spin up-down scattering. The spin
density $n_{-\mu }$ occurs on the right in Eq.~(\ref{rate}) because it is
that of the target particles for the incoming $\mu $ spins. The result is
that 
\begin{equation}
J_{\mu }=-\frac{n_{\mu }}{n_{-\mu }}\frac{\overline{v}k_{B}}{\sigma _{+-}}%
\frac{dT}{dz} .
\end{equation}
If $n_{+}/n_{-}\gg 1$, the heat current $J_{-}$ for down spins is
negligible compared to $J_{+}$ for the up spins and the thermal
conductivity is
\begin{equation}
\kappa =\frac{n_{+}}{n_{-}}\frac{\overline{v}k_{B}}{\sigma _{+-}} .
\end{equation}
For high polarizations ($n_{+}/n_{-}\gg 1$) $\kappa$ can be very large.  
The increase in $\kappa $ and other transport coefficients for 
polarized systems has been predicted 
theoretically\cite{LandL,Meyero,MandM} and also seen 
experimentally.\cite {PolExp} A similar increase also occurs if the 
particles are degenerate.  The idea of a statistical repulsion is 
counterintuitive to this result.  \vspace{0.2in}

\emph{Ferromagnetism}. A very simple mean-field picture of magnetic fluids
is provided by a model in which the particles interact by s-wave scattering
only.\cite{Patria} Thus again there is no up-up or down-down interactions
and the energy of the system of $N_{+}$ up spins and $N_{-}$ down spins is
given by 
\begin{equation}
E=E_{+}+E_{-}+gN_{+}N- ,
\end{equation}
where $E_{\sigma }$ is the total kinetic energy of the $\sigma $ 
spins, which is proportional to $N_{\sigma }^{5/3}$ as in the ideal 
Fermi gas.  The interaction parameter $g$ measures the scattering 
energy between up and down spins.  Two up spins (or two down spins) do 
not ``see'' each other in this model.  If $g>0,$ this model has three 
possible states.  If $g$ is small, then the system favors less kinetic 
energy by having $N_{+}=N_{-}=N/2.$ That is, one has 
antiferromagnetism.  However, if $g$ is large enough, then either all 
the spins are up or all down to minimize the potential energy; 
ferromagnetism results.  The kinetic energy in this case is larger 
than it would be if $N_{\sigma }=N/2,$ but the potential energy is 
zero.  The wave-function antisymmetry between two like spins has made 
them invisible to one another and non-interacting, because their 
minimum separation is greater than the range of the interaction.  The 
idea of a Pauli exchange force would lead one to assume a higher associated 
potential energy (as in Eq.~(\ref{UGpot})), but what actually happens is 
that the \emph{kinetic} energy is raised by having more like spins, 
while at the same time statistics favors \emph{lowering} the potential 
energy.

\section{DISCUSSION}

\label{sec:DISCUSS} Our goals in this paper have been to clarify the idea of
the statistical effects sometimes referred to as ``exchange forces''. 
We believe the
term ``force'' used in this context may mislead students (and even more
advanced workers), who might misinterprete the geometrical effect being
described. 
We have given several examples of
instances where statistical or exchange forces have been invoked to provide
a conceptual explanation of the physics.
We have not introduced any new physics in these examples,
but we have tried to show how a teacher or
writer might provide an alternative interpretation that avoids the
exchange force terminology and thereby arrives at a deeper heuristic
understanding of the physics. Indeed we identified several cases where the
concept of an effective 
statistical force could lead to the opposite of the correct answer.
When a concept has that potential, it is time to replace it.

Our alternative heuristics have taken several forms. We like
Griffiths\cite{Griffiths} wording: ``it is not
really a force at all\ldots it is purely a geometric consequence of the
symmetrization requirement.'' For same-spin fermions, the requirement that
the wave function vanish whenever two particles are at the same position
means that the wave function must have increased curvature, which leads to
an enhanced momentum distribution. Indeed in many cases, the real
statistical effect corresponds to a change in \emph{kinetic} energy (i. e.,
momentum distribution) as in the explanation of the virial pressure or the
white dwarf star, whereas a force picture leads to a change in \emph{%
potential} energy as in the Unlenbeck-Gropper potential of Eq. (\ref{UGpot}%
). Equally helpfully, the geometrical interpretation leads directly to the
changes in average particle separation as compared to distinguishable
particles, with same-spin fermions farther apart on average, which nicely
explains the scattering results where same-spin fermions have a reduced
interaction. 

It is hard to underestimate the importance of the the conceptual element of
physics. Whole introductory courses have been constructed that leave out
much of the mathematical half and concentrate only on the so-called
``conceptual'' side of the subject.\cite{Hewitt} Moreover, we emphasize to
our students that they have not understood a theory until they can 
describe the physics in simple conceptual terms.
Given that emphasis, we offer the following guiding principle regarding
statistical symmetries: ``May the force be \emph{not} with you.''

\begin{acknowledgments}
We would like to thank Dr.\ Terry Schalk for introducing the concern
about Fermi repulsion to us, and Profs.\ John Donoghue and Barry
Holstein for helpful conversations.

\end{acknowledgments}

\newpage

\begin{figure}[tbp]
\centering
\includegraphics[width=6.0in]{UGpot.EPSF}
\caption{Plot of the effective statistical ``interaction'' versus
position. For bosons this function is attractive; for fermions it is
repulsive.}
\label{fig:UGpot}
\end{figure}

\begin{thebibliography}{99}
\bibitem{spinperm}  Eq.~(\ref{Eferm}) can be given a rigorous derivation,%
\cite{Dirac, Herring} in which the factor $\mathbf{\sigma }_{1}\cdot \mathbf{%
\sigma }_{2}$ arises from the spin pair-permutation operator $%
P_{12}^{(S)}=1/2(1+\mathbf{\sigma }_{1}\cdot \mathbf{\sigma }_{2})$ acting
on spin-space functions to ensure correct overall wave function
antisymmetry. The H$_{2}$ problem has analogs in many-body physics in
electron magnetism\cite{Herring,Lieb} and in the nuclear magnetism of solid $%
^{3}$He.\cite{Thouless,McMahan}
\bibitem{surface}  The idea that the energy lowering in the H$_{2}$ molecule
comes from electrons tunneling in a six-dimensional hyperspace from one site
pair to the opposite can be strengthened by expressing $J$ in terms of a
surface integral.\cite{Herring} This surface integral represents a particle
current crossing a hypersurface separating the two localized site
arrangements $(a,b)$ and $(b,a).$
\bibitem{dispersion}  At large separations $r$ between the nuclei of the
atoms, the potential is usually well described as a dispersion or van der
Waal's interaction (due to virtual induced electric dipoles---not included
in the H$_{2}$ discussion), which varies as $-1/r^{6}$. Parameters for
phenominological interatomic potentials are determined from scattering data or
thermal data.\cite{Hirsch}
\bibitem{philess}  
For simplicity, we have assumed that the scattering potential is azimuthally 
symmetric, as it is in so many applications, so that there is no 
dependence on the angle $\phi$.
\end{thebibliography}
\end{document}